\documentclass[fleqn,10pt]{wlscirep}
\title{Multi-GeV electron-positron beam generation from laser-electron scattering}

\author[1,2,*]{Marija Vranic}
\author[2,3]{Ondrej Klimo}
\author[2]{Georg Korn}
\author[2]{Stefan Weber}
\affil[1]{GoLP/Instituto de Plasmas e Fus\~ao Nuclear, Instituto Superior T\'ecnico, Universidade de Lisboa, 1049-001 Lisbon, Portugal}
\affil[2]{ Institute of Physics of the CAS, ELI-Beamlines Project, Na Slovance 2, Prague, 182 21, Czech Republic}
\affil[3]{ Faculty of Nuclear Sciences and Physical Engineering, Czech Technical University in Prague, Brehova 7, Prague, 115 19, Czech Republic}

\affil[*]{marija.vranic@ist.utl.pt}

\begin{abstract}
The new generation of laser facilities is expected to deliver short (10 fs - 100 fs) laser pulses with 10 - 100 PW of peak power.
This opens an opportunity to study matter at extreme intensities in the laboratory and provides access to new physics.
Here we propose to scatter GeV-class electron beams from laser-plasma accelerators with a multi-PW laser at normal incidence. 
In this configuration, one can both create and accelerate electron-positron pairs. 
The new particles are generated in the laser focus and gain relativistic momentum in the direction of laser propagation. 
Short focal length is an advantage, as it allows the particles to be ejected from the focal region with a net energy gain in vacuum.
Electron-positron beams obtained in this setup have a low divergence, are quasi-neutral and spatially separated from the initial electron beam. 
The pairs attain multi-GeV energies which are not limited by the maximum energy of the initial electron beam. 
We present an analytical model for the expected energy cutoff, supported by 2D and 3D particle-in-cell simulations. 
The experimental implications, such as the sensitivity to temporal synchronisation and laser duration is assessed to 
provide guidance for the future experiments. 
\end{abstract}

\begin{document}

\flushbottom
\maketitle
\thispagestyle{empty}

\section*{Introduction}

Generating abundance of antimatter in laboratory is of great importance both for fundamental science and potential applications. Several scenarios have been envisaged to create plasma clouds (or beams) with approximately equal number of electrons and positrons, dense enough to exhibit collective behavior. The aim is to study their self-consistent dynamics in conditions that match the ones the pair plasmas experience in space. Electron-positron pairs populate the magnetospheres of pulsars, and are believed to participate in the formation of gamma-ray bursts \cite{GRB_review}. It is therefore vital to study the collective processes under extreme intensities, which may play a key role in the global dynamics of pulsar magnetospheres. Apart from laboratory astrophysics, $e^+e^-$ pairs have a number of other prospective applications that span from tests of fundamental symmetries in the laws of physics and studies of antimatter gravity to the characterisation of materials \cite{Review_positrons_traps}. 

It is thus not surprising that a strong effort is placed towards generating electron-positron plasmas using high-power laser technology. One of the most promising all-optical ways to accomplish this is via the non-linear Breit-Wheeler process \cite{BW_paper}.  In Breit-Wheeler (BW) pair production, a particle emits a high-energy photon, that can possibly decay into an electron-positron pair in an intense electromagnetic background. Such a strong background field can nowadays be produced by an intense laser. A proof-of-principle experiment was performed at SLAC \cite{E144slac1}, where BW pairs were produced in a collision of a 46.6 GeV electron beam from a linear accelerator with a green laser at the intensity of $10^{18}~\mathrm{W/cm^2}$. 
Despite the resounding success of the experiment, a very small fraction of electrons composing the beam gave rise to a hard $\gamma$-ray that would eventually decay into a pair. 
The yield was too low to be used for a pair plasma study. 
%Even though this experiment was exciting from the fundamental point of view, only a tiny fraction of electrons emitted a photon that later created a pair, and one could not use this setup for studying the dynamics of electron-positron plasmas. 
On the other hand, several other laser experiments  \cite{CGahn_exppairsBH1,CGahn_exppairsBH2, HChen_exppairs1, HChen_exppairs2, cowan1999, Chen_POP_2015, Chen_recent_exp, Sarri_ep_exp, Sarri_older_BH_exp} use Bethe-Heitler process for pair generation \cite{BH_paper}. This is similar to the BW pair production, with one difference: a strong field background is provided by the proximity of a high-Z nucleus instead of a high-power laser pulse. Positron creation is possible also without using lasers  \cite{Positron_traps, Stellarator_ep}. However, a challenge in such schemes is positron accumulation and storage, in addition to the difficulty of recombining them with the electrons to later create an electron-positron flow. 

Multi-PW laser facilities currently under construction  \cite{ELI_WhiteBook, QED_at_ELI, ELI_NP_2016, WEBER_P3} will access the regime of prolific pair production \cite{Bell_Kirk_2lasers, RevModPhysdipiazza}. During scattering with such intense laser beams, relativistic electrons emit high-frequency radiation through Compton scattering. Recent milestone all-optical experiments scattered electrons with lasers at 180 degrees, and demonstrated the potential of the state-of-the-art laser technology to generate x-rays and $\gamma$-rays  \cite{MalkaNature, Sarri_compton_exp, Chen_compton_exp, Powers_compton_exp, Khrennikov_compton_exp}. A recent review on laser-wakefield acceleration-based light sources can be found in Ref \cite{Review_Albert_Thomas} and the most recent results on multiphoton Thompson scattering in Ref \cite{Umstadter_multiphoton_t}. All these experiments were performed below the radiation reaction dominated regime, because the overall energy radiated by the interacting electrons was small compared with the initial electron energy. By using more intense laser pulses ($I \sim 10^{22}~ \mathrm{W/cm^2}$) or more energetic electron beams, we will soon be able to convert a large fraction of the electron energy into radiation and access the regime of quantum radiation reaction  \cite{Ridgers_quantumRRnew, Piazza_qed_energyspread, Piazza_QRR_FD, Dinu_harvey, Yoffe_evolution, Vranic_quantumRR}. This is expected in the next few years, as ~4 GeV electron beams have already been obtained using a 16 J laser \cite{Leemans_4gev} and the next generation of facilities is aiming to achieve laser intensities $I > 10^{23}~ \mathrm{W/cm^2}$. In such extreme conditions, the energetic photons produced in the scattering can decay into electron-positron pairs  \cite{Lobet_Davoine_pairs}. 
\begin{figure*}
	\includegraphics[width=1.0\textwidth]{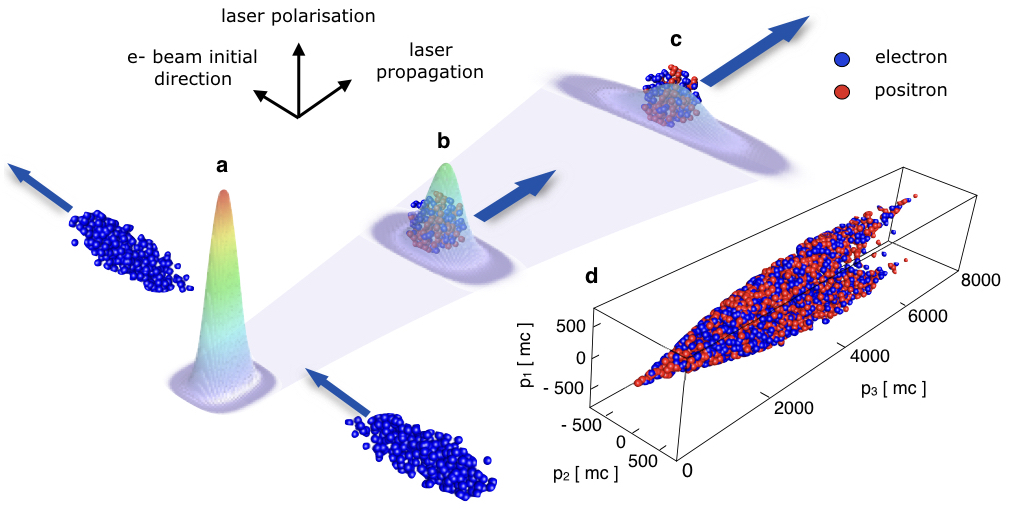}
	\caption{Setup. (\textbf{a}) Perpendicularly moving electron beam interacts with the laser at the focus and creates new pairs; (\textbf{b}) Some electrons and positrons obtain a momentum component parallel to the laser and start getting accelerated; (\textbf{c}) The laser defocuses shutting down the interaction; this leaves the particles with the net energy gain from the laser. (\textbf{d}) A fraction of the accelerated electrons and positrons distributed within the momentum space. } 
	\label{setup_pic}
\end{figure*}
Here we propose a configuration that allows to both create and accelerate an electron-positron beam. An intense laser interacts with a relativistic electron beam at 90 degrees of incidence (setup is illustrated in Fig. \ref{setup_pic}). 
The pair production efficiency here is slightly lower than in a head-on collision. 
However, in a head-on collision the energy cutoff of the electron-positron beam is limited to the initial energy of the interacting electrons, while at 90 degrees this is not the case.
At 90 degrees of incidence, if generated with a low energy, new particles can be trapped and accelerated in the laser propagation direction. 
If the created particles are very energetic, they continue emitting hard photons to further feed the pair creation. 
Once their energy is low enough to be trapped, they rapidly develop a momentum component parallel to the laser propagation direction that supresses the quantum interaction. 
The creation and the acceleration phase are therefore decoupled. Due to the laser defocusing, the trapped particles remain in the laser field only a fraction of a full oscillation cycle. 
This limits the maximum energy they can attain, but allows for a net energy transfer in vacuum that would otherwise be impossible. 
We have developed a predictive analytical model for the energy cutoff of the electron-positron flow generated in the electron-laser scattering. 
Our theory is supported by full-scale 2D and 3D particle-in-cell (PIC) simulations, where the quantum processes are modelled via an additional Monte-Carlo module. 
We show that this setup produces a neutral electron-positron flow that can reach multi-GeV energies. The flow has a divergence of about $\sim$ 30 mrad. 
A distinguishing aspect of this scheme is to produce at extreme intensities an equal number of electrons and positrons that can be separated from the initial electron beam. 
The original electrons are, in fact, reflected before entering the region of the highest laser intensity where most pairs are created. 
As a result, the pairs and the earlier reflected electrons move in slightly different directions and can be collected separately. 

\section*{Results}

\subsection*{Pair generation and quantum parameters.}

Laser intensity, electron energy and their relative angle of incidence determine whether classical or quantum processes dominate the laser-electron interaction.
One way to quantitatively distinguish between the two regimes is through a Lorentz-invariant dimensionless parameter $\chi_e$, that is formally defined as \cite{Ritus_thesis}
\begin{equation}\label{chie_def}
	\chi_e=\sqrt{(p_\mu F^{\mu \nu})^2}/(E_c m c)
\end{equation}
Here, $E_c=m^2c^3/(\hbar c)$ is the critical field of electrodynamics that can perform a work of $mc^2$ over the Compton length (and can spontaneously create electron-positron pairs in vacuum), $F^{\mu \nu}$ is the electromagnetic tensor, $m, \ p_\mu$ are the electron mass and 4-momentum, and $c$ is the speed of light. 
Physically, parameter $\chi_e$ represents the ratio of the electric field of the laser to $E_{c}$ when observed from the particle's rest frame. 
For $\chi_e\ll 1$, the interaction between the laser and the particle is purely classical. 
For $\chi_e\gg1$, the interaction is quantum-dominated and one expects to create a vast number of electron-positron pairs through a two-step Breit-Wheeler process. 

Breit-Wheeler pair production is mediated by energetic photon emission through Compton scattering. 
A single photon can carry a large fraction of the emitting particle's energy.
Later on, in the presence of an intense background field, a fraction of such photons decays into electron-positron pairs. 
Analogously with $\chi_e$, one can define a photon quantum parameter $\chi_\gamma=\hbar\sqrt{(k_\mu F^{\mu \nu})^2}/(E_c m c)$, where $\chi_\gamma$ of the emitted photon is always smaller than $\chi_e$ of the electron where the emission takes place.
The photon $\chi_\gamma$ can later grow if the photon moves to a region of a more intense electromagnetic field.  
There is no sharp threshold for Breit-Wheeler pair production, but the rate at which the process occurs is exponentially suppressed for $\chi_\gamma\ll1$ where the number of pairs scales as $\sim \exp (-8\chi_\mathrm\gamma/3)$. 
Hence, electrons in the transition between the classical and the quantum interaction regime ($0.1<\chi_e<1$) are not expected to generate many photons that eventually decay into pairs. 
However, these particles can convert a significant fraction of their energy into radiation, either in the form of a single energetic photon or through emission of many low-energy photons.  

We have mentioned previously that the electron $\chi_e$ in a background field depends on its direction of motion. 
From Eq. (\ref{chie_def}), it is clear that different scattering angles can result in a different $\chi_e$ for otherwise the same electron beam and laser parameters.
This is a consequence of the Einstein's theory of relativity: the laser field amplitude has a certain value in the laboratory frame, but this value is higher or lower in another reference frame.
For the pair production to occur, the laser field is required to be on the order of $E_c$ in the electron's rest frame.
One can, therefore, exploit a relativistic velocity of the electron beam to increase the effective laser intensity as perceived by the particles. 
The most favourable configuration for obtaining high values of $\chi_e$ is a scattering where the laser and the electron beam are anti-parallel.   
A simplified form of Eq. \eqref{chie_def} for a counter-propagating electron is  $\chi_e\simeq2\gamma_0 a_0 \times \hbar \omega_0/(mc^2)$, where  $\gamma_0$ is the Lorentz factor, $\omega_0$ is the laser frequency and $a_0=0.85 \sqrt{I[10^{18}\ \mathrm{W/cm^2}]} \ \lambda_0[\mu\mathrm{m}]$ is the normalized vector potential of the laser field. 
For a co-propagating particle this value is $\chi_e\simeq (a_0/(2\gamma_0)) \times \hbar \omega_0/(mc^2)$, while for the interaction at 90 degrees $\chi_e\simeq \gamma_0 a_0 \times \hbar \omega_0/(mc^2)$. For example if a 1 GeV particle beam counter-propagates with a laser field of intensity of $10^{23}~\mathrm{W/cm^2}$, $\chi_e\approx2$, for co-propagation $\chi_e\approx1.4\times10^{-7}$ and for interaction at 90 degrees $\chi_e\approx1$. Therefore, a co-propagating particle interacts classically with the laser, while a counter propagating particle is in a fully quantum regime of interaction. The famous SLAC-E144 experiment \cite{E144slac1} was performed with $\chi_e=0.3$.

Energetic photons (e.g. gamma-rays) are emitted by relativistic particles in the direction of particle propagation (more accurately, within a cone whose pitch angle scales as $\theta\sim1/\gamma$). The same is true when a photon decays into a pair of particles: due to the 4-momentum conservation, the new particles share almost the same propagation direction with the decaying photon. From that, it follows that if the electron bunch interacts with the laser at 90 degrees of incidence, the photons are also emitted at 90 degrees, and the pairs they produce also move in the same direction. This process continues until some particles are left with the energy small enough such that the laser rapidly introduces a parallel momentum higher than the perpendicular component and suppresses the quantum interaction. 

As we will see later, the particles with the lowest initial energies have a potential to attain the highest energy in the end. 
A conservative estimate for the lowest energy available at the time of creation through the Breit-Wheeler process corresponds to the electrons and positrons with $\chi_e \simeq 0.1$. 
One can obtain such particles either by creating them directly, or through classical or quantum radiation reaction.
Below $\chi_e = 0.1$, the radiation reaction is too weak to induce a significant energy loss on a particle within a short interaction time (on the order of tens of femtoseconds for a particle interacting at 90 degrees of incidence with a laser focused on a few $\mu \mathrm{m}$ spot size). 
One can therefore not guarantee the presence of particles with initial $\chi_e<0.1$, even though having such particles in the system is possible.

\subsection*{Particle acceleration and pulse defocusing. Prediction for the $e^+e^-$ energy cutoff. }

We will now discuss how the particles can be accelerated in the field of the same laser that has provided the intense background for their creation.
Lawson-Woodward theorem states that an unbounded plane wave in vacuum cannot provide net acceleration over an infinite interaction time \cite{Lawson_LW, Woodward_LW}.  
For the acceleration to occur, at least one condition of the Lawson-Woodward theorem must be violated.  
This can be achieved, for example, by introducing a background gas in the interaction region, or using optical components \cite{Esarey_PRE_1995} or oblique gratings  \cite{palmer_grating_linac} to limit the interaction distance.
Vacuum acceleration of particles has been proposed previously in the field of two crossed laser beams \cite{Esarey_PRE_1995, Sprangle_1996}, or with higher-order Gaussian beams \cite{Esarey_PRE_1995}. 
Here, we exploit the fact that the laser is a tightly focused Gaussian beam to accelerate particles. 
This concept was proposed previuosly for lasers focused beyond the diffraction limit, where electrons were injected externally \cite{Keitel_Salamin_tight_focusing}.
However, in our case, a large fraction of new particles is generated in the laser focus, the region of the highest electromagnetic energy density. 
These particles can have an arbitrary initial phase and are created with a relativistic momentum (in the direction of propagation of the decaying photon). 
The newly born electrons and positrons start oscillating within the laser field.  
As the laser is spatially bounded, some particles do not have time to complete a full oscillation cycle before they leave the region of the intense field where they are created.  
They can leave the intense field region either laterally (by moving away from the laser axis) or longitudinally (by propagating further than the Rayleigh length). 
The limited interaction time due to the spatial bounds of the high-intensity region allows particles to obtain net acceleration in vacuum. 

It is important to note that the pair generation process is decoupled from the particle forward acceleration (in the direction of laser propagation). 
Once particles acquire a momentum component parallel to the laser axis, this naturally suppresses their quantum parameter ($\chi_e\propto a_0/\gamma$ for co-propagating particles).  
By changing the direction of the momentum vector, quantum effects become quenched and classical relativistic interaction with the laser takes over.   
One can, therefore, apply the Landau \& Lifshitz equation of motion \cite{LLbook} for the particle forward acceleration.

Classical particle motion in a plane wave can be described analytically, even in cases when the radiation reaction is of relevance. 
There exists an exact solution for a particle born at an arbitrary phase of the plane wave \cite{Piazza_solutionLL}. 
We use this solution as a base, to which we later introduce corrections for the acceleration in a focused laser pulse. 
Below we give a flavor of the main ideas and crucial steps, while a more precise description of the procedure is found in the Supplementary Material \cite{suppl_m}.    

Energy gain in a plane electromagnetic wave depends on the initial and the final phase of the particle within the wave, laser intensity and the particle initial energy. 
We assume that the initial particle momentum is perpendicular to both, the laser propagation direction and the electric field of the laser. 
Under these conditions, one can estimate the maximum attainable instantaneous energy to be 
\begin{equation}\label{emax}
	\frac{\mathcal{E}_\mathrm{max}}{mc^2}=\frac{2 a_0^2}{\gamma_0}.
\end{equation}
The maximum possible energy $\mathcal{E}_\mathrm{max}$ is inversely proportional to the initial Lorentz factor $\gamma_0$ because $\gamma_0$ affects the phase slippage between the particle and the wave during forward acceleration. 
Within the same wave, a particle with a lower initial energy has a potential to obtain a higher final energy. 
However, the actual acceleration achieved also depends on the initial and the final phase within the wave. 
One obtains $\mathcal{E}_\mathrm{max}$ just for the most favourable case, for which the interaction stops exactly when $I(\phi)=4$ (c.f. \cite{suppl_m}). 
This is an idealistic view, as the interaction in realistic conditions does not have a sharp, but a gradual shutdown. 
The value $\mathcal{E}_\mathrm{max}$ can, however, serve as an estimate for a beam energy cutoff. 
After a certain time of interaction $t_\text{acc}$, the $e^+e^-$ beam is expected to have a continous spectrum with all particles in the energy range  $0<\mathcal{E}<\mathcal{E}_\mathrm{max}$. Here, $t_\text{acc}$ is the time required for a particle in best conditions for acceleration to reach the maximum energy $\mathcal{E}_\mathrm{max}$.

Following the analysis from the previous section, we assume the lowest available initial energy $\gamma_0$ in the system corresponds to the electron quantum parameter $\chi_e=0.1$. 
Using the simplified $\chi_e$ definition for interaction at normal incidence, we can relate the lowest $\gamma_0$ with the laser intensity: 
\begin{equation}\label{gam0_a0}
	\gamma_0 \simeq 5\times10^4/a_0.
\end{equation} 
The maximum attainable energy then depends only on the available laser intensity: 
$\mathcal{E}_\mathrm{max}=4 \times 10^{-5}a_0^3~mc^2$. 
A natural conclusion follows: the more intense the laser is, the more energetic $e^+e^-$ pairs one can obtain. 
However, as we mentioned before, the acceleration is not immediate. 
The acceleration length required to reach the maximum energy in the field of a plane wave is given by $l_\mathrm{pwa}=3 a_0^2\lambda_0/(8\gamma_0^2)$. 
For example, if we have a laser at $I=10^{23}\ \mathrm{W/cm^2}$ and $\lambda_0=1~\mu\mathrm{m}$, a maximum attainable energy would be $\mathcal{E}_\mathrm{max}=5.7\times10^4\ mc^2$. 
But, $80\ \mu\mathrm{m}$ of propagation would be required to obtain that energy. 
It is clear that even the most powerful laser facilities currently under construction \cite{ELI_WhiteBook} will not be able to provide such a long acceleration distance. 
The intensity on the order of $10^{23}\ \mathrm{W/cm^2}$ will be achieved by focusing the laser to a tiny waist $W_0$, which will result in a quick laser defocusing characterized by the Rayleigh length $R_L=\pi W_0^2/\lambda_0$. 
For example, a laser focused to a spotsize of $W_0=3\ \mu\mathrm{m}$, has the Rayleigh length $R_L=28\ \mu\mathrm{m} $, which is much smaller than the required acceleration distance $l_\text{pwa}=80~\mu\mathrm{m}$. 
The particles will not spend enough time in the vicinity of the peak of the laser field to reach $\mathcal{E}_\mathrm{max}=5.7\times10^4\ mc^2$.

However, if we know the focusing parameters of the laser beam, we can estimate the correction to the expected energy cutoff for the $e^+e^-$ beam. 
To this end, we have studied test particle motion in 3D, with the focused laser pulse field defined as in Ref. \cite{Mora_PRE_1998}.
The detailed analysis presented in Supplemental Material \cite{suppl_m} shows that there are several cases to consider.  If $l_\mathrm{pwa}\ll R_L$, the acceleration is completed before the laser defocuses significantly. 
We therefore assume that for $R_L>10\ l_\mathrm{pwa}$, the cutoff energy of the beam is unaltered: $\mathcal{E}_\mathrm{cutoff}=\mathcal{E}_\mathrm{max}$.
If $l_\mathrm{pwa}\gtrsim R_L$, then the field intensity varies during the acceleration time and the realistic energy cutoff, in this case, is lower than the theoretical maximum $\mathcal{E}_\mathrm{max}$. 
If $R_L <l_\mathrm{pwa}$, one can approximate the decrease in cutoff energy as (c.f. Suppl. material \cite{suppl_m}): 
\begin{equation}\label{cutoff}
	\mathcal{E}_\mathrm{cutoff}=\mathcal{E}_\mathrm{max} \sqrt{0.5~ R_L/l_\mathrm{pwa}},
\end{equation}
where the ratio between the Rayleigh length and the acceleration distance is given by 
$R_L/l_\mathrm{pwa}= (8\pi/3) ( \gamma_0/a_0)^2 (W_0/\lambda_0)^2$. 
It follows from here that for arbitrarily low initial Lorentz factor $\gamma_0$, the maximum attainable energy given by Eq. (\ref{cutoff}) depends only on the laser parameters and for a laser with $\lambda_0=1~\mu$m can be expressed as: 
\begin{equation}\label{cutoff1}
	\mathcal{E}_\text{cutoff}\ \mathrm{[MeV]}\simeq2a_0\ W_0~[\mu\mathrm{m}].
\end{equation}

We reach the conclusion that the particle acceleration has two possible limits. 
One is a plane wave acceleration limit dependent on the laser intensity and initial Lorentz factor given by Eq. (\ref{emax}), and the other one that depends only on the laser parameters given by  Eq. (\ref{cutoff1}). Depending on the energy of the particles born within the laser field, the lowest of the two limits determines the actual cutoff.

\subsection*{QED-PIC simulations of particle generation and acceleration. Effect of 3D geometry and long propagation distance on acceleration of self-created $e^+e^-$ beam.} 

\begin{figure*}
	\begin{center}
		\includegraphics[width=0.7\textwidth]{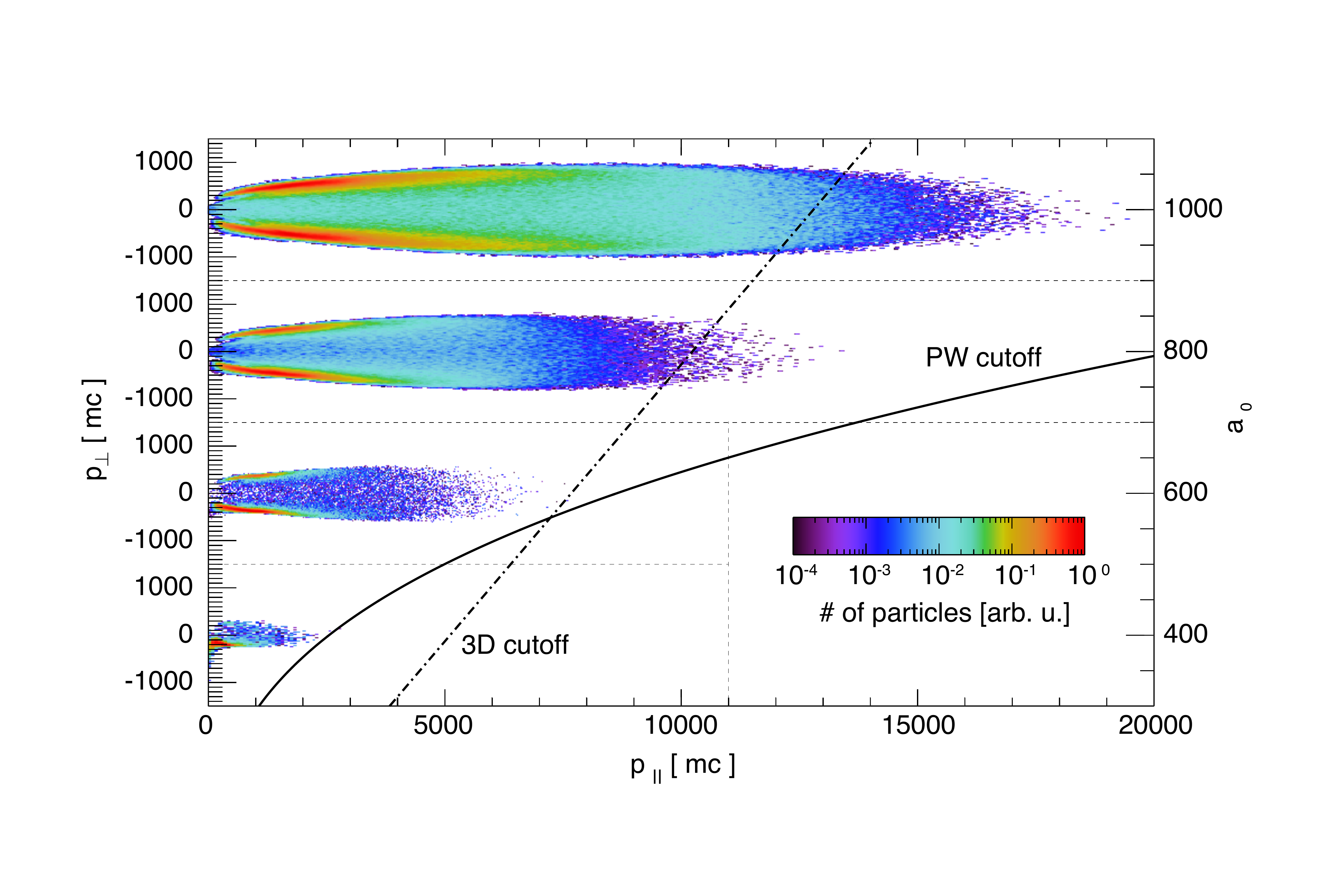}
		\caption{ Momentum spaces of the generated positrons from 2D QED simulations with lasers of $a_0\ = $ 400, 600, 800 and 1000 where $p_{||}$ is the momentum in the direction of laser propagation, and $p_\perp$ is the momentum in the initial electron beam propagation direction. The laser duration is 150 fs, and the waist $W_0=3.2\ \mu\mathrm{m}$. Full line denotes the plane wave interaction energy cutoff determined by Eqs. (\ref{emax}) and (\ref{gam0_a0}); dash-dot line represents the cutoff given by Eq. (\ref{cutoff1}) expected in 3D due to laser defocusing. } 
		\label{cutoff_theory_pic}
	\end{center}
\end{figure*}

\begin{figure*}
	\includegraphics[width=1.0\textwidth]{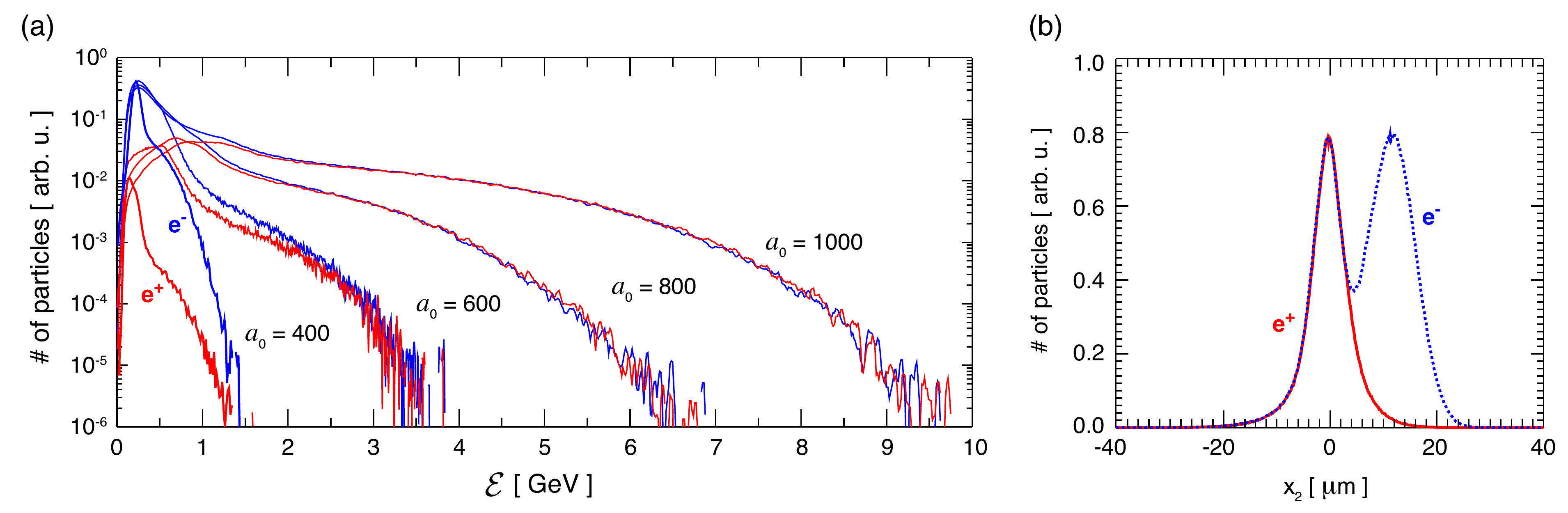}
	\caption{ (\textbf{a}) Electron and positron spectra for 150 fs  lasers at several  intensities. The number of electrons and positrons in the beam with the energy above 2 GeV is approximately equal. (\textbf{b}) Electron and positron density integrated over the main laser propragation direction for $a_0=1000$. Pairs are spatially separated from the initial electron beam that cannot penetrate into the highest intensity region. The $e+e-$ bunch and the $e^-$ beam continue propagating away from one another.} 
	\label{spectrum}
\end{figure*}

The ideas presented in the previous section can be confirmed by full-scale QED-PIC simulations, that model the $e^+e^-$ pair creation and subsequent acceleration beyond the Rayleigh length. 
We have performed a set of simulations in 2D geometry, for lasers with $\lambda_0=1~\mu\mathrm{m}$, $W_0=3.2~\mu\mathrm{m}$ and duration of $t=150~\text{fs}$.  
Figure \ref{cutoff_theory_pic} shows a momentum space in the realistic aspect ratio for the self-created $e^+e^-$ beam at several different laser intensities. 
The lines denote the maximum parallel momentum consistent with the analytical prediction for the energy cutoff given by Eqs. (\ref{emax}) and (\ref{cutoff1}). 
Two-dimensional QED-PIC simulations show the newly created particles get accelerated while trapped within the intense part of the laser field. 
The initial direction of propagation of the electron beam is not imprinted in the high-energy section of the pair momentum space (as can be confirmed in Fig.  \ref{cutoff_theory_pic}). 
The reason is that the particles with lowest initial energies populate the high-end of the energy spectrum. 
They obtain transverse momentum due to the wavefront curvature (particles are accelerated in the direction of the local $\vec{k}$ vector) that is much higher than their initial momentum.
The particles with low $p_\parallel$ in Fig. \ref{cutoff_theory_pic} are created with high $p_\perp$ and can have a preferential $p_\perp$ towards the initial electron propagation direction (hence a slight asymmetry in Fig. \ref{cutoff_theory_pic}, especially for low values of $a_0$). 
The beam remains, however, with a low divergence of $\sim$30 mrad. 
Energies up to 9 GeV were obtained with $a_0=1000$ (see Fig. \ref{spectrum}a). 
Above 2 GeV, the spectrum has an equal number of positrons and electrons. 
The differences between the electron and positron spectra in the low-energy section are due to the initial electron beam (the spectrum contains all the particles in the simulation box). 
The electrons from the initial beam can continue propagating forward, having lost a portion of their energy to photons (as depicted in Fig. \ref{setup_pic}). 
Another option is that the electrons lose enough energy to get reflected by the laser field backwards, or get trapped and accelerated forward the same way as the pairs. 
In our case, for intensities $a_0>600$, most electrons lose energy so rapidly that they never get to interact with the peak of the field - they get reflected sooner.
The result is a spatial separation between the pairs that were created in the peak field and the initial electrons.    
One can therefore collect a neutral electron-positron beam on the detector by applying an appropriate aperture. 
This can be confirmed in Fig. \ref{spectrum}b, where the $e^+e^-$ bunch has already separated spatially from the electron beam (the two density peaks propagate away from one another).  

A natural question to arise is whether the newly-created particles are accelerated the same way when they are allowed to move in all three spatial directions. 
As it is known, the electromagnetic fields decay differently in 2D and 3D geometry, and it is vital to perform 3D simulations to assess the viability of future experiments.
In addition, full 3D field structure could affect particle trapping and the effective interaction time. 

Unfortunately, it is not possible to perform full-scale 3D QED simulations for a sufficiently long propagation distance to obtain a satisfactory estimate of the $e^+e^-$ properties far away from the laser focus. 
But, to add fidelity to our previous conclusions, we can make use of the fact that the pair generation is temporally decoupled with the acceleration process. 
This is not strictly true for the entire beam, but on the level of individual particles, it is a reasonable enough assumption.
If a particle is created with a sufficiently low energy, we assume it does not contribute to further pair generation and its trajectory can be treated classically. 
If it is created with a large momentum (that provides $\chi_e\gtrsim1$), then this particle has a significant probability to radiate an energetic photon and contribute in creating a new pair. 
However, after this process has been repeated several times, the initial and all subsequently generated particles have the energy low enough to start the classical interaction.

\begin{figure*}
	\includegraphics[width=1.0\textwidth]{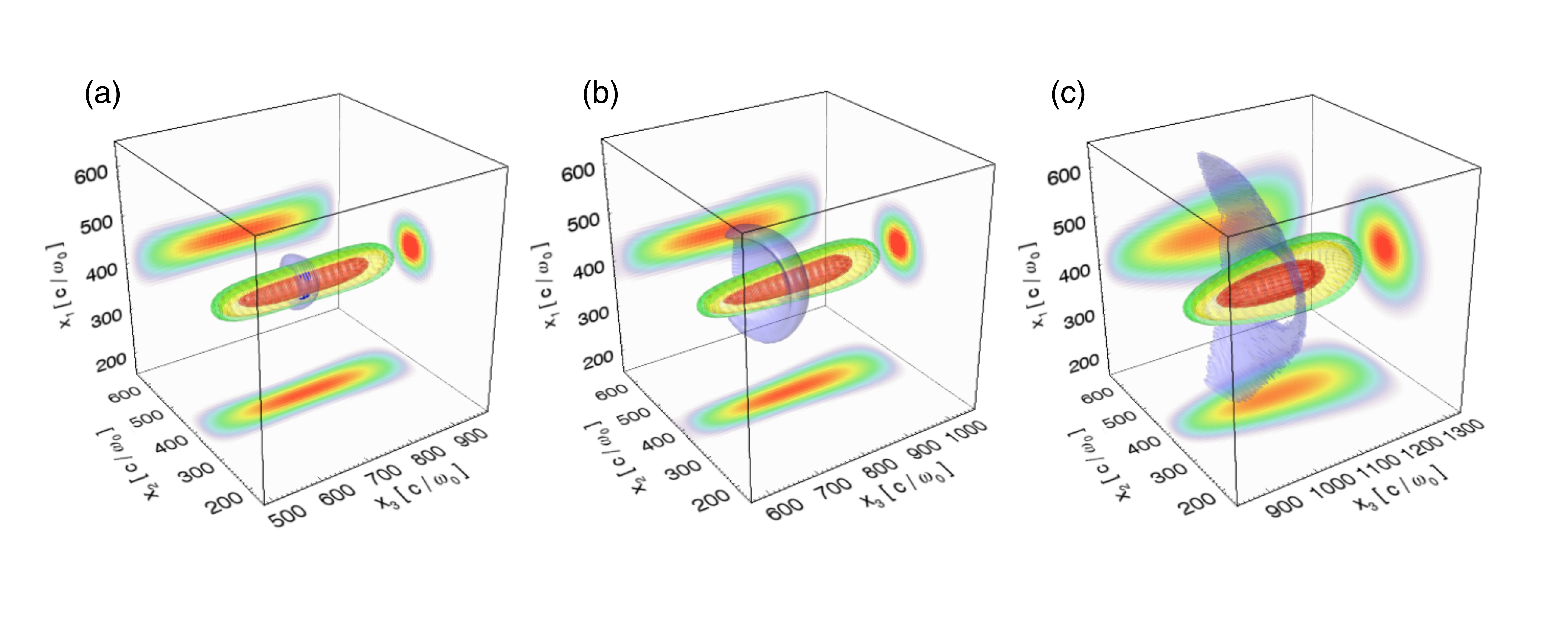}
	\caption{3D simulations of long $e^+e^-$ beam propagation within a 150 fs laser beam with $W_0=3.2~\mu\mathrm{m}$. The $e^+e^-$ beam is initially at the laser focus, with an initial momentum at 90 degrees from the laser propagation direction. Average Lorentz factor is $<\gamma>=250$ with a large energy spread of  50\% to mimic the distribution of the self-created particles. Laser peak $a_0=600$. The red isosurfaces correspond to the local field amplitude of $a_0=100$, the yellow ones correspond to $a_0=20$ while the green isosurfaces correspond to $a_0=10$. (\textbf{a}) $t=32\ \omega_0^{-1}$ (\textbf{b}) $t=112\ \omega_0^{-1}$ (\textbf{c}) $t=352\ \omega_0^{-1}$.  } 
	\label{3d_plots}
\end{figure*}

\begin{figure*}
	\includegraphics[width=1.0\textwidth]{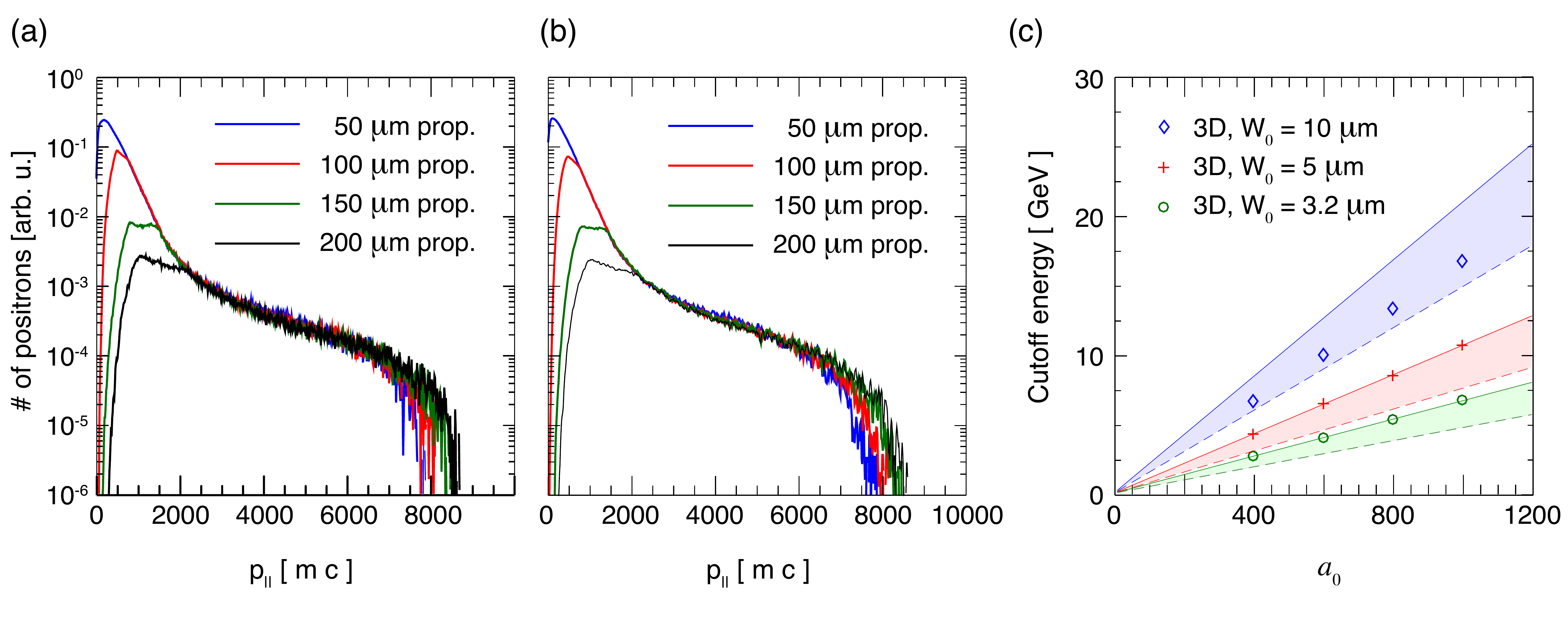}
	\caption{Parallel momentum distribution for positrons after propagating to a certain distance from the focal plane.  (\textbf{a}) $\tau=150~\mathrm{fs}$ (same as Fig. \ref{3d_plots}), (\textbf{b}) $\tau=30~\mathrm{fs}$. (\textbf{c}) Cutoff energy as a function of laser intensity $a_0$ and waist $W_0$. Full lines represent $\mathcal{E}_\text{cutoff}$ given by Eq. \ref{cutoff1}, and dashed lines represent $\mathcal{E}_\text{cutoff}/\sqrt{2}$. The shown results are for $\tau=150~\mathrm{fs}$.}
	\label{3d_plots_line}
\end{figure*}

Laser defocusing and the beam evolution in the simulation are illustrated in Fig \ref{3d_plots}. 
Here, the laser duration is 150 fs, and the figure shows three distinct instants of time.
One striking fact is that the initial propagation direction is not a preferential direction of lateral escape for the particles. 
This is well illustrated in Fig. \ref{3d_plots}b, where all directions perpendicular to the laser propagation seem to be equivalent. 
The particle density isosurface shows a certain curvature.
It is clear that the ponderomotive force due to the slowly-varying laser envelope could not have created such a curvature at the central point of the laser. 
However, as particles move away from the focal plane, the wavefront of the field they experience becomes curved. 
This means that their acceleration in the laser direction of propagation (perpendicular to the wavefront) can introduce curvature in the final particle density distribution. 
Figure \ref{3d_plots}c shows that after some time, the particle density is not cylindrically symmetric but elongated in the laser polarisation direction. 
This is due to the transverse momentum obtained within the laser field. 

We can now evaluate how long it takes for the particle momentum/energy distribution to stabilize. 
Figure \ref{3d_plots_line}a shows the momentum in laser propagation directions for particles initialized within a 150 fs laser. 
The distributions displayed correspond to the times after the laser central point has propagated 50 $\mu$m, 100 $\mu$m, 150 $\mu$m and 200 $\mu$m away from the focal plane. 
One sees that the higher energy end of the particle momentum distribution does not change significantly after the propagation beyond 50 $\mu$m. 
This is not surprising, because the Rayleigh length for this laser is $R_L\simeq32~ \mu$m, and most of the acceleration occurs close to the focus. 
The differences in the number of particles in the low-energy range are due to the particles leaving the simulation box. 
Figure \ref{3d_plots_line}b shows the parallel momentum distribution of an identical particle beam like in panel a), but in the field of a shorter (30 fs) laser.  
The cutoff is of the same order and the conclusions about the propagation distance are similar as before.
This is not surprising, as the particles that reach the highest energies are nearly phase-locked with the laser for a while (the interaction lasts a fraction of the cycle in the frame co-moving with the laser), and then escape laterally. 
The laser temporal envelope cannot have a strong effect on their acceleration in such conditions. 

Figure \ref{3d_plots_line}c shows the energy cutoff from 3D simulations as a function of $a_0$ and $W_0$. 
The coloured regions show where we expect the cutoff if we have low enough $\gamma_0$ in the system. 
Upper bound for $\mathcal{E}_\text{cutoff}$ is given by Eq. (\ref{cutoff1}) and shown with full lines. 
Lower bound (dashed lines) is given by $\mathcal{E}_\text{cutoff}/\sqrt{2}$, which represents the maximum energy attainable by a particle born exactly at the centre of the pulse (c. f. Supplemental Material \cite{suppl_m}). 
For $W_0=3.2~\mu\mathrm{m}$ and $W_0=5~\mu\mathrm{m}$, the maximum energy obtained in the 3D simulations is exactly given by Eq. (\ref{cutoff1}). 
For $W_0=10~\mu\mathrm{m}$, the values are slightly below the maximum possible value, but within the expected region. 
One should, in principle, be able to achieve the maximum given by Eq. (\ref{cutoff1}) for $W_0=10~\mu\mathrm{m}$ by varying the initial conditions. 

It is noteworthy that for $a_0>600$, the cutoff prescribed by Eq.  (\ref{cutoff1}) is lower than what we obtain in 2D QED simulations, while for $a_0=400$ it is higher than that. 
Due to the slower field fallout in 2D, the effective interaction time is longer, so the 2D simulation cutoff is expected to be higher than the 3D cutoff predicted by Eq. (\ref{cutoff1}). 
However, this is not neccesarily true at lower laser intensities, because there may be too few particles generated with low values of $\gamma_0$. 
In that case, Eq. (\ref{emax}) with the lowest available $\gamma_0$ within the laser focus determines the maximum asymptotic energy in the system (as can be confirmed from Fig. \ref{cutoff_theory_pic}).

\subsection*{Number of pairs in the $e^+e^-$ spectrum; the importance of laser duration and temporal synchronization} 

\begin{figure*}
	\begin{center}
	\includegraphics[width=0.8\textwidth]{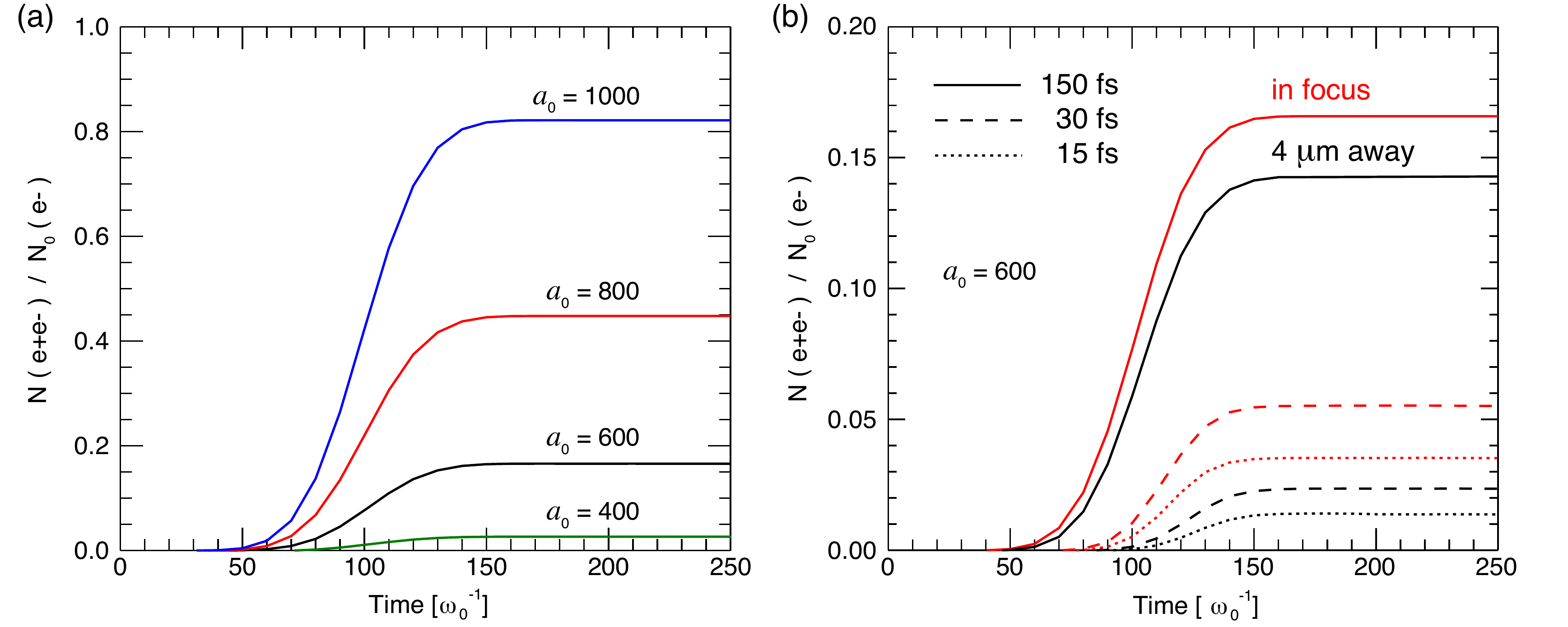}
	\caption{Number of generated pairs $N (e^+e^-)$ divided by the initial number of electrons $N_0 (e^-) $ as a function of time.  (\textbf{a}) We vary the laser intensity for a fixed laser duration of 150 fs. (\textbf{b}) We keep the laser intensity fixed while varying the duration. Red lines show the interaction in focus, with perfect temporal synchronization (the centre of the interacting $e^-$ beam overlaps with the laser perfectly in the moment when the highest intensity section of the laser reaches the focus). Black lines are for a non-ideal temporal synchronization (the $e^-$ beam is in the laser focus at the instant of time when the peak of the laser intensity is 4 $\mu$m away from the focal plane).  } 
	\label{number_pairs}
	\end{center}
\end{figure*}

As we have seen in the previous section, the energy spectrum cutoff given by Eqs. (\ref{emax}) or (\ref{cutoff1}) depends on the laser intensity, and does not significantly change if one uses a shorter pulse. 
Nevertheless, the laser duration is important for the total number of pairs one can create in this setup. 
To illustrate this, we now return to the QED-PIC simulations. 
Figure \ref{number_pairs} shows the number of pairs generated per initial electron as a function of time. 
With a $150\  \mathrm{fs}$ laser at $a_0=800$ and $W_0=3.2\ \mu\mathrm{m}$, one can generate a pair per 45\% of initial 0.5 GeV electrons.  
As expected, the number of generated pairs is higher for higher laser intensities. 
Figure \ref{number_pairs} b) shows that for a fixed laser intensity $a_0=600$, shorter lasers generate fewer pairs than the longer ones. 
This is obviously expected for counter-propagating scattering configurations because the interaction time is longer. 
However, here the interaction is at 90 degrees, where the time of a single electron within the laser field is determined by the width of the focal spot, rather than the laser duration. 
But, the electron beam has spatial dimensions, and not all the electrons interact with the laser at its peak intensity. 
If a laser has a longer temporal envelope, the intensity around the focus drops slower. 
This gives a chance for more electrons to interact with high laser intensities. 
Similar ideas apply to the fact that the number of pairs is more sensitive to the temporal synchronization for shorter laser pulses, which is also shown in Fig. \ref{number_pairs}b.   
It is, therefore, better to use a longer laser pulse to perform this experiment if one has access to the same laser intensity. 

Even though the acceleration itself is not that affected by the pulse duration, the initial distribution of pairs at creation time is different (a longer pulse creates more particles overall and has more particles sampling the low-energy range).
This results in a slightly different shape of the energy spectrum, and in some cases a different cutoff (please note that the conclusion in the previous section about the same cutoff for different pulse durations was drawn from the same initial pair distribution).
%For $a_0=400$ 15fs pulse, less than 0.01\% of particles above 1 GeV, for 30fs 0.04, for 150 fs 0.14.
This can be illustrated at the $e^+e^-$ spectrum after the acceleration is completed for a laser intensity corresponding to $a_0=600$. 
For a 150 fs pulse, 22\% of the total pairs have forward kinetic energy above 1 GeV, while 15 \% of pairs are in such conditions for a 30 fs pulse and only 10\% for a 15 fs pulse. 
With a 150 fs laser, for $a_0=800$,  59 \% of pairs are in the part of the spectrum above 1 GeV and 25 \% above 2 GeV.
For $a_0=1000$, 78 \% particles are above 1 GeV and 55 \% above 2 GeV.
The measurements are taken after 50 $\mu$m of propagation from the focus beyond which the distribution does not significantly change (as confirmed by Fig. \ref{3d_plots_line}).

\section*{Discussion}

A new generation of laser facilities will be able to create and accelerate electron-positron beams to multi-GeV energies. 
The setup we propose in this manuscript will allow generating for the first time a neutral, highly-collimated relativistic flow of electron-positron pairs in the laboratory.
The total number of pairs is tunable by varying the interacting electron energy. 
We have studied in detail the case starting with a 1 GeV electron beam, where producing one positron per interacting electron requires a laser intensity above $10^{23}~\mathrm{W/cm^2}$ ($a_0\simeq1000$). 
However, for the same laser waist $W_0=3~\mu$m, a 3 GeV electron beam can generate 1.8 pairs per electron with a $a_0=600$, and a 4 GeV beam produces 1.2 pairs per electron already with $a_0=400$. 
Bearing in mind that 4 GeV electron beams are current state-of-the-art \cite{Leemans_4gev}, and the laser-wakefield electron beams envisaged for the near future will have charge  $\gtrsim 10$ nC \cite{Martins_LWFA}, it will be possible to generate a neutral pair plasma tens of skin depths wide in each direction. 
These electron-positron pairs can be accelerated to energies well above 1 GeV, and above the energy of the initial scattering electrons. 
Generating such beams is of fundamental interest as a step towards future positron accelerators.
The proposed setup can be used as an all-optical source directly, or to generate seed particles that can be accelerated further using other acceleration techniques. 

\section*{Methods}

\subsection*{Particle-in-cell} codes use Maxwell's equations discretized on a grid to advance the electromagnetic fields. Charge density is sampled by a list of macro-particles moving in the 6D phase space. The electromagnetic fields are interpolated from the grid to the particle positions, and their momenta are updated according to the Lorentz force (or Landau \& Lifshitz equation of motion if radiation reaction is relevant for the simulation). The currents produced by moving charges are deposited on the grid nodes, where the electromagnetic fields are advanced in the next iteration. Laser pulses are initialized assuming the paraxial approximation. Detailed implementation techniques for OSIRIS can be found in Ref. \cite{OSIRIS}.

\subsection*{QED PIC module} A Monte-Carlo module that accounts for the photon emission and Breit-Wheeler pair produciton is implemented as an addition to the standard PIC loop of OSIRIS. Energetic photons are initialised as an additional particle species. The differential probability rate of photon emission for nonlinear Compton scattering is given by $d^2P/(dt d\chi_\gamma)=\alpha mc^2/(\sqrt{3}\pi \hbar\gamma\chi_e)\times [\tilde{\eta}\  K_{2/3}(\tilde{\chi})-\int_{\tilde{\chi}}^{\infty}dx\  K_{1/3}(x)]$, where $\alpha=e^2/(\hbar c)$ is the fine structure constant, $\gamma$ is the Lorentz factor of the emitting electron, $K$ is the modified Bessel funciton of the second kind, $\tilde{\eta}=(\chi_e^2+(\chi_e-\chi_\gamma)^2)/(\chi_e(\chi_e-\chi_\gamma))$ and $\tilde{\chi}=2\chi_\gamma/(3\chi_e(\chi_e-\chi_\gamma))$. The differential rate of pair production by a photon in a background electromagnetic field is given by 
$d^2P/(dt d\chi_e)=\alpha m^2 c^4/(\sqrt{3}\pi \hbar^2\omega\chi_\gamma)\times[\tilde{\eta}\ K_{2/3}(\tilde{\chi})+\int_{\tilde{\chi}}^{\infty}dx\ K_{1/3}(x)]$, where $\hbar\omega$ is the energy of the decaying photon,
$\tilde{\eta}=(\chi_e^2+(\chi_\gamma-\chi_e)^2)/(\chi_e(\chi_\gamma-\chi_e))$ and $\tilde{\chi}=2\chi_\gamma/(3\chi_e(\chi_\gamma-\chi_e))$. These emission rates are found in Refs \cite{pair_rate1,pair_rate2,pair_rate3}. OSIRIS QED has been used previously in Refs. \cite{ThomasQED, Thomas_POP_2016, MV_ppcf2016, Vranic_quantumRR}. A similar method for incorporating BW pair production is used in several other codes  \cite{Elkina_rot, Ridgers_solid, Nerush_laserlimit, Bell_Kirk_MC, Gonoskov2015review, Gonoskov2014, Gremillet, lobet,  Basmakhov}.

\subsection*{Details on modelling the acceleration stage} To describe the classical acceleration, one can initialize an electron-positron beam already within the laser field, moving at an angle of 90 degrees compared to the laser direction of propagation. 
To mimic the conditions of the self-created electron-positron beam at the laser focus, one should initialize the energy distribution as close as possible to the one we observe in 2D simulations that include QED processes.
The advantage of this approach is that it is now possible to perform the simulations with the standard PIC algorithm, that is faster and does not require a timestep as small as in the QED-PIC simulations.   
This allows evaluating how the beam evolves within the laser field in 3D for long propagation distances. 
The approximations of this approach are that particles are assumed to be created all at the same time and the acceleration starts immediately. 
Also, all values for the initial phase $\phi_0$ within the laser are taken with equal probability, whereas in QED the particles are more likely to be created (or radiate) in the regions where the electric field is close to its peak value. 
But, this still samples an arbitrary initial phase and therefore all possible individual particle outcomes. 
These particles can be accelerated forward and escape laterally from the laser field. 
Even though due to the above approximations we do not expect the spectrum to be exactly the same as in QED simulations, we can estimate how long it takes for the acceleration to be completed and for the spectrum to stabilize (by allowing the particles to move far enough from the intense field region).  

\subsection*{Simulation parameters} All 2D simulations are performed with OSIRIS QED. The box is $160\ \mu\mathrm{m}$ long and $80 \ \mu\mathrm{m}$ wide, resolved with $10000\times2500$ cells. Total simulation time is 210 fs, with a timestep of $dt=0.53\  \mathrm{as}$. Laser normalised vector potential is varied between $a_0=400$ and $a_0=1000$ and the laser is polarised in the $x_1$ direction. Three  laser durations are used: $15\ \mathrm{fs}$,  $30\ \mathrm{fs}$ and $150\ \mathrm{fs} $. The interacting electron beam initially moves perpendicularly to both, the laser propagation and polarisation direction. The direction of the electron beam propagation is within the focal plane of the laser. The electron beam is initialised as a Gaussian density distribution,  $3.2\ \mu\mathrm{m}$ wide at FWHM with a peak density of $n=0.01~n_c$. The central kinetic energy of the beam is 0.5, 1 or 2 GeV. The 3D simulations of long $e^+e^-$ propagation are performed with the standard OSIRIS algorithm using the moving window and taking radiation reaction into account. The simulation box is $160\  \mu\mathrm{m}$ long and $127\ \mu\mathrm{m} \times 127\ \mu \mathrm{m}$ wide
resolved with $5040\times2040\times2040$ cells. Total simulation time is 3.190 ps with a timestep of $0.08 \ \mathrm{fs}$. The laser is initialised at the focus with the peak normalised potential varied between $a_0=400$ and $a_0=1000$, and waist sizes of $W_0=3.2\ \mu\mathrm{m},\ 5 \mu\mathrm{m}$ and $10\ \mu\mathrm{m}$. Two laser durations considered were 30 fs and 150 fs. The electron-positron beam overlaps with the laser at the focus and has an initial energy of 125 MeV (a Gaussian distribution with about 50\% energy spread) to mimic the distribution of the Breit-Wheeler pairs created in the 2D QED simulations. 

%\bibliography{paper_scattering}

\section*{Acknowledgements}

This work is supported by the grants: HIFI - CZ.02.1.01/0.0/0.0/15\_003/0000449,  ELITAS - CZ.02.1.01/0.0/0.0/16\_013/0001793, Czech Science Foundation Project No. 15-02964S.  and LaserLab Europe IV-GA. N. 654148 (H2020-INFRAIA-2014-2015). The simulations were performed at clusters Eclipse (ELI Beamlines) and Salomon (IT4Innovations). The authors would like to acknowledge the OSIRIS Consortium, consisting of UCLA and IST (Lisbon, Portugal) for providing access to the OSIRIS framework. We thank Dr T. Grismayer for fruitful discussions. 

\section*{Author contributions statement}

M. V. developed the analytical model, performed the simulations and analyzed the data. 
O. K. and S. W. gave valuable comments throughout the research project.  
G. K. provided overall project support.
M. V. wrote the manuscript with input from O. K. and S. W. All authors read the manuscript. 

\section*{Additional information}

\textbf{Competing financial interests} The authors declare that they have no
competing financial interests. 

\clearpage

\section*{Suplemental Material}
\vspace{10pt}

\subsection*{A general solution for particle motion in a plane wave with arbitrary initial conditions}

This section sumarises the analytical solution for relativistic particle motion in a plane electromagnetic wave with an arbitrary initial phase presented in Ref. \cite{Piazza_solutionLL}.  Particle energy lost due to the classical radiation reaction is included in the model. 

The wave amplitude is taken to be constant, and there is no feedback on the wave itself. According to Ref \cite{Piazza_solutionLL}, relativistic 4-velocity within a linearly polarized wave can be expressed as:
\begin{equation}\label{diPiazza1}
	u^\mu(\phi)=\frac{1}{h(\phi)} \left\{ u_0^\mu+ \frac{1}{\rho_0}[h^2(\phi)-1] n^\mu - \frac{1}{\rho_0}I_1(\phi)\frac{e f_1^ {\mu\nu}}{m} u_{0,\nu}+ \frac{1}{2\rho_0}\xi_1^2 I_1^2(\phi)n^\mu \right\},  
\end{equation}
where $ \xi_1^2=\xi^2=(e E_0/m\omega_0)^2$,  $f_1^{\mu\nu}=n^\mu a_1^\nu-n^\nu a_1^\mu$,  $a_1^ \mu = (0, a_0, 0, 0)^T$ is the normalized vector potential and  $n^\mu= (1,\vec{n})$. 
Here, $\vec{n}$ is a  unit vector parallel to the laser direction of motion $\vec{n}\ ||\ u_3$ 
and $\rho_0\equiv (n u_0)$. 
Einstein notation was used (a repeated Greek index is equivalent to a summation over all dimensions and $(ab)\equiv a^\mu b_\mu$).
Functions $I_1(\phi)$ and $h(\phi)$ carry the imprint of the radiation reaction. They are expressed as:   
\begin{equation}\label{func_rr}
	\begin{split}
		h(\phi) & =1+\frac{1}{3}\alpha \eta_0 \xi^2 \left[\omega_0(\phi-\phi_0)-\sin(\omega_0\phi)\cos(\omega_0 \phi)+\sin(\omega_0\phi_0)\cos(\omega_0\phi_0) \right] \\
		I_1(\phi) & = \cos(\omega_0\phi)h(\phi) - \cos(\omega_0\phi_0)-\frac{2}{3}\alpha\eta_0 \left(\sin(\omega_0\phi)-\sin(\omega_0\phi_0)\right) \times \\
		& \times \left[1+\frac{\xi^2}{3} \left(\sin^2(\omega_0\phi)+\sin(\omega_0\phi)\sin(\omega_0\phi_0) + \sin^2(\omega_0\phi_0)) \right)  \right],
	\end{split}
\end{equation}
where $\phi_0$ denotes the initial phase, $\phi$ current phase of the particle within the field of the electromagnetic wave, $\alpha$ is the fine structure constant, and $\eta_0= (k_0u_0)/m$ is the wave frequency measured in the electron rest frame in units of the electron mass $m$.  Other quantities denote elementary charge $e$ and electromagnetic field amplitude $E_0$. 
Without radiation reaction, Eqs. \eqref{func_rr} reduce to 
\begin{equation}\label{diPiazza3}
	h(\phi)=1, \quad  I_1(\phi) = \cos(\omega_0\phi)-\cos(\omega_0 \phi_0).
\end{equation}
The equations (\ref{diPiazza1})-(\ref{diPiazza3}) are presented in natural units, as in Ref. \cite{Piazza_solutionLL}. In those units $\hbar=c=1$, and the fine-structure constant is defined by $\alpha=e^2/\hbar c\simeq1/137 $. We now switch to the dimensionless units more common in laser-plasma interactions. Here, all the quantities are normalized to the laser frequency $\omega_0$, such that $t\rightarrow t \omega_0$, $p\rightarrow p/mc$, $\mathcal{E}\rightarrow \mathcal{E}/mc^2$, $E\rightarrow eE/mc\omega_0$ and $B\rightarrow eB/mc\omega_0$. In these units, $\omega_0=1$, $\lambda_0=2\pi$, $c=1$, mass is normalized to the electron mass $m$ and charge is normalized to the elementary charge $e$. The particle equations of motion by component for the case without radiation reaction starting at arbitrary laser phase with an arbitrary initial momentum can be written as: 
\begin{equation} \label{eqs_motion}
	\begin{split}
		& \gamma(\phi)  =  \gamma_0 + a_0 \frac{p_{0,1}} {\gamma_0-p_{0,3}} I_1(\phi) + \frac{1}{2(\gamma_0-p_{0,3})}a_0^2 I_1(\phi)^2\\
		& p_1(\phi)  = p_{0,1} + a_0  I_1(\phi) \\
		& p_2(\phi)  = p_{0,2} \\
		& p_3(\phi)  = p_{0,3} + a_0 \frac{p_{0,1}} {\gamma_0-p_{0,3}} I_1(\phi) + \frac{1}{2(\gamma_0-p_{0,3})}a_0^2 I_1(\phi)^2  
	\end{split}
\end{equation}
We can compute the spatial displacement of the particle as a function of the phase $\phi$. We have $x_3=\int_{0}^{t} p_3(t)/\gamma(t)dt=\int_{\phi_0}^{\phi} p_3(\phi)/(\gamma(\phi)-p_3(\phi))d\phi$, where we used a relation between the phase and time $d\phi=dt (\gamma-p_3)/\gamma$ to perform the change of variables.
The denominator is an integral of motion, such that $\gamma(\phi)-p_3(\phi)=\gamma_0-p_{0,3}$. The spatial displacement in the laser direction of motion is then given by: \\
\begin{equation}\label{x3}
	\begin{split}
		x_3= &\frac{p_{0,3}}{\gamma_0-p_{0,3}}(\phi-\phi_0)+ \frac{a_0 p_{0,1}}{(\gamma_0-p_{0,3})^2} \left[ \sin\phi -\sin \phi_0 - (\phi-\phi_0) \cos \phi_0 \right]\\
		+& \frac{a_0^2}{2(\gamma_0-p_{0,3})^2} \left[ (\phi-\phi_0)\left(\cos^2\phi_0+\frac{1}{2}\right)-2 \cos\phi_0 \sin\phi + \frac{1}{4} \sin(2\phi) + \frac{3}{4}\sin(2\phi_0)\right]
	\end{split}
\end{equation}
The standard result for a particle starting at rest can be retrieved by taking  $\gamma_0=1 $ and $p_{0,1}=p_{0,2}=p_{0,3}=0$. However, our particles are not created at rest. Electron-positron pairs are created by scattering an LWFA electron bunch with an intense laser at normal incidence. We can assume as a first approximation that $p_{0,1}=p_{0,3}=0$, but $p_{0,2}\neq 0$, i.e. the particle initial velocity is perpendicular to both the laser polarization direction $x_1$ and to the laser propagation direction $x_3$. 

\subsection*{Maximum energy and plane wave acceleration length for particles initially at 90 degrees}

For a particle born at 90 degrees of incidence to the laser axis with an initial Lorentz factor $\gamma_0$, maximum attainable energy according to Eqs. (\ref{eqs_motion}) is given by $\mathcal{E}_{\text{max}}=2a_0^2/\gamma_0$. This is achieved when $I_1(\phi)^2$ is at the maximum. For simplicity, let us consider a particle with $\phi_0=0$, as an example that allows to obtain that maximum. Here, the expression for longitudinal displacement given by Eq. (\ref{x3}) can be simplified:
\begin{equation}\label{simple_x3}
	x_3= \frac{a_0^2}{2\gamma_0^2}\left(\frac{3}{2}\phi - 2 \sin\phi+\frac{1}{4}\sin(2\phi)\right).
\end{equation}
The maximum energy $\mathcal{E}_\text{max}$ is reached for $\phi=\pi$, after the particle has traveled a distance of $l_{\text{pwa}}=3a_0^2\pi/(4\gamma_0^2)$ or $l_{\text{pwa}}=\lambda_0\times 3a_0^2/(8\gamma_0^2)$. 
We define this distance as the plane wave acceleration length. If our particle would start at a different $\phi_0$ in otherwise identical conditions, its maximum energy can be smaller or equal to $\mathcal{E}_\text{max}$. 
For particles born at normal incidence, the particles with the lowest initial $\gamma_0$ are the ones that can attain the highest energy in the interaction with the plane wave (because  $\mathcal{E}_\text{max}\propto \gamma_0^{-1}$). 
However, one should note here that the higher energy gain is possible due to the slower dephasing, and the plane wave acceleration length is reversely proportional to the square of the initial energy $l_\text{pwa}\propto\gamma_0^{-2}$. 
Such particles, therefore, propagate a longer distance before reaching the $\mathcal{E}_\text{max}$. 

\subsection*{Correction of the maximum energy due to the pulse focusing} 

The maximum energy and the acceleration length in the previous section are valid for a plane wave interaction. 
However, the lasers are not plane waves, and the laser de-focusing affects the acceleration. 
Particles are expected to attain an energy cutoff similar to $\mathcal{E}_\text{max}$ in a focused laser only if $R_L \gg l_{\text{pwa}}$, where $R_L=\pi W_0^2/\lambda_0$ is the Rayleigh range,  $W_0$ and $\lambda_0$ are the laser waist and the wavelength respectively. 
If $R_L\simeq l_\text{pwa}$, we require an estimate on how the laser de-focusing affects the acceleration. 
To this end, we perform a series of test particle simulations, where the particles are initialized with the most favourable phase for the highest energy gain within the shortest propagation distance. 
The test particles are born at the laser central axis, in focus, at $\phi_0=0$ of the wave.  
The 3D analytical expression for the fields of the focused laser pulse is taken from Ref. \cite{Mora_PRE_1998}.
We varied the laser intensity, and the initial particle energy is chosen according to $\gamma_0=5\times10^4/a_0 $ to be consistent with the lowest available energy expected in the system. 
\begin{figure*}
	\begin{center}
		\includegraphics[width=0.5\textwidth]{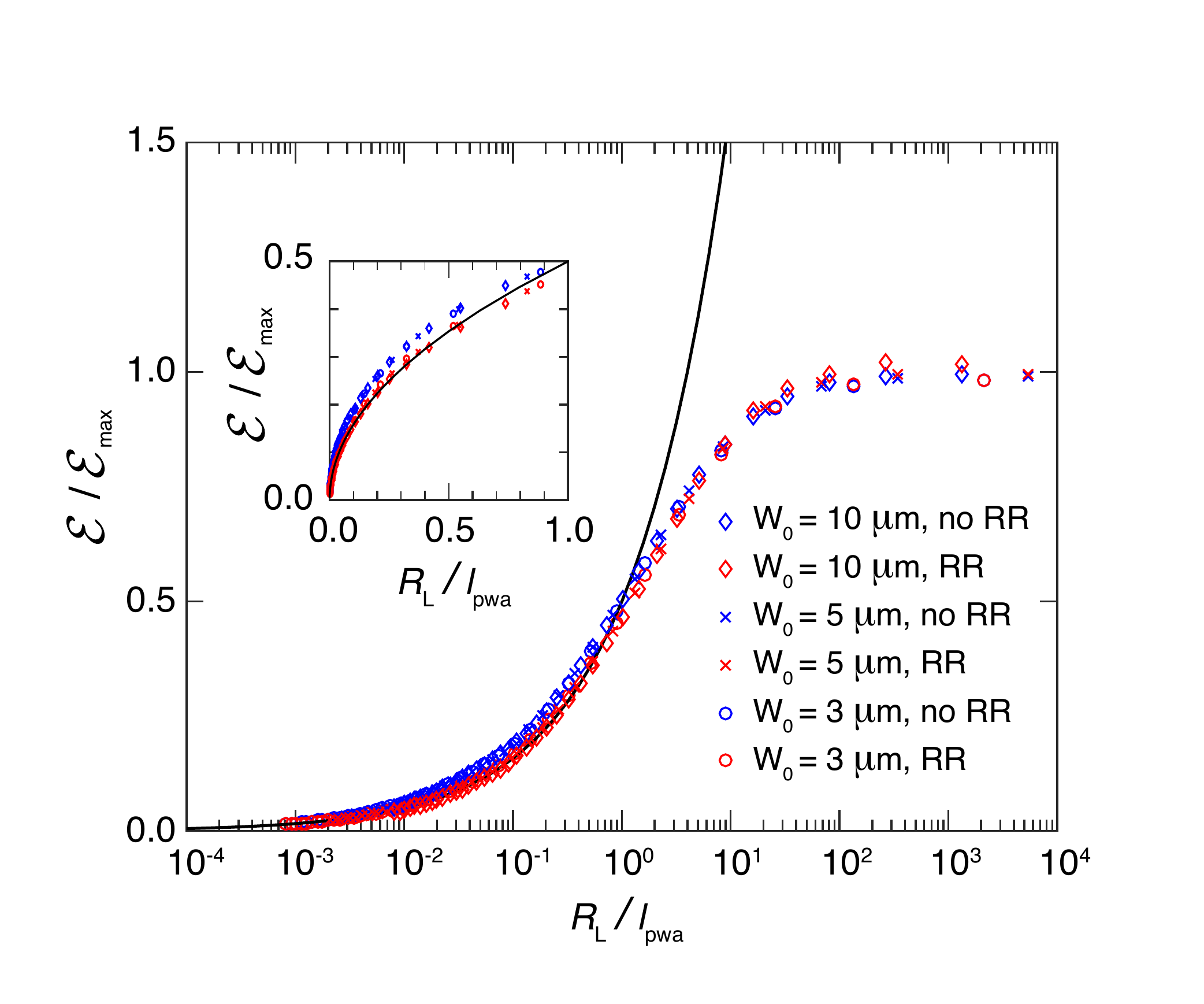}
		\caption{Ratio of the maximum attainable energy for a focused laser $\mathcal{E}$ and the maximum attainable energy for the same laser intensity in a plane wave $\mathcal{E}_\text{max}$ as a fuction of the ratio of the Rayleigh range $R_L$ to the acceleration length $l_\text{pwa}$. For $0.05<R_L/l_\text{pwa}<1$, one can approximate the correction of maximum energy due to the pulse focusing using the Eq. (\ref{ratio}). } 
		\label{fraction}
	\end{center}
\end{figure*}

The results displayed in Fig. \ref{fraction} correspond to a range of laser intensities between $a_0=100$ and $a_0=4000$ and three different waist sizes: $W_0=3.2\ \mu\mathrm{m}$, $W_0=5\ \mu\mathrm{m}$ and $W_0=10\ \mu\mathrm{m}$. 
For every example, the maximum energy attained by the particle born in the laser focus with $\phi_0=0$ is divided by the maximum energy such a particle could attain in a plane wave. 
An interesting point to note here is that the differences between the case with and without radiation reaction are not strongly pronounced. 
This is consistent with the fact that a quickly acquired parallel component of the momentum reduces the $\chi_e$ and the radiation reaction stops being relevant for the particle dynamics.
Figure \ref{fraction} indicates that the ratio $\mathcal{E}/\mathcal{E}_\text{max}$ is a function of $R_L/l_\text{pwa}$ and for $R_L/l_\text{pwa}<1$ this can be approximated as: 
\begin{equation}\label{ratio}
	\frac{\mathcal{E}}{\mathcal{E}_\text{max}}=0.5 \sqrt{ \frac{R_L}{l_\text{pwa}} } 
\end{equation}
For $R_L/l_\text{pwa}>10$, we consider the condition $R_L\gg l_\text{pwa}$ satisfied and $\mathcal{E}\simeq\mathcal{E}_\text{max}$. 
For $l_\text{pwa}<R_L< 10~ l_\text{pwa}$, Eq. (\ref{ratio}) is not a good estimate, due to its value being on the same order of $\mathcal{E}_\text{max}$. For  $l_\text{pwa}<R_L< 10~ l_\text{pwa}$, the maximum energy takes values between $\mathcal{E}_\text{max}/2$ and $\mathcal{E}_\text{max}$.

Equation (\ref{ratio}) shows the defocusing correction for the maximum energy of a particle born in the very centre of the laser field. 
We can assume that for an arbitrary particle born anywhere in the laser field (with the same $\gamma_0$), the effective interaction length cannot be more than twice the value for the particle born in the centre. As the effective interaction length is proportional to $R_L$, we should not expect to obtain particles with energies higher than
\begin{equation}\label{cutoff_si}
	\frac{\mathcal{E}}{\mathcal{E}_\text{max}}=0.5 \sqrt{ \frac{2R_L}{l_\text{pwa}} } .
\end{equation}
Full-scale 3D simulations in the main manuscript are in very good agreement with the predictions for the cutoff of the particle energy spectrum given by Eq. (\ref{cutoff_si}).

\end{document}